\documentclass[preprint,aps,prd,12pt,preprintnumbers,nofootinbib,superscriptaddress]{revtex4-2}
\usepackage{xcolor}
\usepackage{braket}
\usepackage{amssymb,amsmath,amsfonts,comment,latexsym,graphicx,epstopdf,float}
\usepackage{color}
\usepackage{fancyvrb}
\usepackage{chngcntr}
\usepackage{etoolbox}
\usepackage{ulem}
\usepackage{array}
\usepackage{bbold}
\usepackage{subfigure}
\usepackage{slashed}
\usepackage{multirow}

\usepackage[colorlinks=true,citecolor=blue,linkcolor=red,filecolor=purple,urlcolor=magenta]{hyperref}

\usepackage{float}

\newcommand{\be}{\begin{equation}}
\newcommand{\ee}{\end{equation}}
\newcommand{\ba}{\begin{eqnarray}}
\newcommand{\ea}{\end{eqnarray}}

\newcommand{\grts}{\raise.3ex\hbox{$>$\kern-.75em\lower1ex\hbox{$\sim$}}}
\newcommand{\lets}{\raise.3ex\hbox{$<$\kern-.75em\lower1ex\hbox{$\sim$}}}

\usepackage{color}
\usepackage{amssymb,amsmath,amsfonts}
\usepackage{epstopdf} 
\usepackage{braket}

\usepackage{autobreak}

\begin{document}
%
%
\title{\vspace*{0.5in} 
Improved unimodular black holes with self-consistent renormalization scale identification
\vskip 0.1in}
\author{Christopher D. Carone}\email[]{cdcaro@wm.edu}
\affiliation{High Energy Theory Group, Department of Physics, William \& Mary, Williamsburg, VA 23187-8795, USA}
\date{June 3, 2026}
%
%
\begin{abstract}
We consider the renormalization group improvement of unimodular black hole metrics with an identification of the renormalization scale as a function of the radial coordinate that is self-consistent in that it depends on parameters of the improved metric.  Considering identifications that are motivated by minimality and dimensional analysis, we arrive at a number of non-singular unimodular black hole metrics.  We determine the black hole mass gaps and note the qualitative similarities to other non-singular black hole metrics that have been discussed in the literature.
\end{abstract}

\maketitle
\newpage

\section{Introduction} \label{sec:intro}
Non-singular black hole metrics have garnered significant attention in the literature given the belief that a quantum theory of gravity should lead to the resolution of the singularities at the origin~\cite{Bardeen:1968,Hayward:2005gi,Dymnikova:1992ux,Simpson:2018tsi,Frolov:2021vbg,Bonanno:2000ep}.   In the absence of such a theory, phenomenological metrics have been proposed and studied, including the well-known examples of  Bardeen~\cite{Bardeen:1968} and Hayward~\cite{Hayward:2005gi}, in hopes that they might capture the qualitative features that might follow from an underlying quantum theory of gravity.  Phenomenological metrics generally involve ad hoc modifications of exact solutions to Einstein's equations, so that they do not represent solutions themselves.    The additional terms that are generated when these modified metrics are substituted into the Einstein tensor are often interpreted ex post facto as an effective quantum gravitational contribution to the energy-momentum tensor~\cite{Hayward:2005gi,Frolov:2016pav}.

A somewhat less ad hoc method of creating regular black hole metrics is through the process of renormalization group improvement~\cite{Bonanno:2000ep,Bonanno:2006eu,Koch:2013owa,Eichhorn:2022bgu}.  In this approach, the nonperturbative running of the gravitational constant is determined via the functional renormalization group~\cite{Polchinski:1983gv,Wetterich:1992yh,Morris:1993qb,Reuter:1996cp,Berges:2000ew}, and the running constant $G(k)$ is incorporated into the metric, with the renormalization scale $k$ identified with a function of the spatial coordinates.  As noted in Ref.~\cite{Bonanno:2000ep}, a similar approach applied in quantum electrodynamics correctly reproduces the Uehling potential, with the identification of the renormalization scale $k \sim 1/r$.  Applied to gravity, renormalization group improvement provides a systematic way of introducing a new functional dependence on $r$ in a spherically symmetry metric compared to the classical solution. However, the results depend on the scale-setting prescription used to identify the renormalization scale $k$ with a physical length or curvature scale~\cite{Bonanno:2000ep,Koch:2013owa,Gonzalez:2015upa,Pawlowski:2018swz,Platania:2019kyx}.

The renormalization group improvement of unimodular black holes has been considered previously in Ref.~\cite{Torres:2017ygl}, with the identification of the renormalization scale associated with quantities calculated using the classical metric.   The metric for a Schwarzschild black hole can be written in unimodular form by change of coordinates, but equivalence at the classical level does not persist in the quantum theory, where the restriction on the metric limits the allowed graviton field configurations that are included in the generating functional for correlation functions~\cite{Eichhorn:2013xr}.    Unimodular gravity is interesting since the cosmological constant appears as a constant of integration rather than as a fundamental Lagrangian parameter and is not renormalized~\cite{Alvarez:2015sba}.  This alleviates the fine-tuning problem associated with the cosmological constant, reducing it to an issue of technical naturalness, which is generally considered to be an improvement.

In this work, we revisit the issue of renormalization group improved unimodular black holes and consider other identifications of the renormalization scale.  We are motivated by the conjecture that this identification might be better formulated in terms of the quantum corrected metric rather than the classical one.   Platania~\cite{Platania:2019kyx} has investigated this possibility through an iterative process in which a renormalization scale that is determined by a quantum corrected gravitational constant is used to generate an improved value which can be used subsequently to determine a new renormalization scale.   When that process converges, one obtains a definition of the renormalization scale that is consistent with the corrected form of the metric.   The cases we consider here effectively focus on this self-consistent limit point, which is independent of how this iterative sequence is initially seeded, leading to non-singular metrics that have not been associated previously with unimodular black holes.  Our results are relatively simple and may be useful as benchmarks for the further study of unimodular black holes, even if one prefers other approaches to the identification of the renormalization scale.

Our paper is organized as follows.  In Sec.~\ref{sec:ug}, we briefly review unimodular gravity and the running of the gravitational constant in that context.   In Sec.~\ref{sec:models}, we present two models that follow from two different types of self-consistent identifications of the renormalization scale. In Sec.~\ref{sec:conc}, we discuss possible interpretations of our scale identifications, suggest future work, and summarize our conclusions.

 \section{Unimodular Gravity} \label{sec:ug}
 
 In unimodular gravity, the determinant of the metric $g \equiv -\det g_{\mu\nu}$ is not dynamical and one sets $\sqrt{g}$ to a constant; we will assume
 \begin{equation}
 \sqrt{g} =1 \,\,\, .
 \end{equation}
 At the classical level, this constraint can be achieved via a coordinate transformation.  For example, the spherically symmetric, static metric
\begin{equation}
ds^2 = -f(r) \, dt^2+ \frac{1}{f(r)}dr^2 + r^2 d\Omega^2
\end{equation}
can be put in unimodular form via the change of variables
\begin{equation}
\rho = \frac{r^3}{3} \,\,\,\,\, \mbox{ and } \,\,\,\,\, x=-\cos\theta \, ,
\end{equation}
yielding~\cite{Torres:2017ygl}
\begin{equation}
ds^2 = -f(\rho)\, dt^2+\frac{1}{(3 \,\rho)^{4/3}} \frac{d\rho^2}{f(\rho)} + (3\, \rho)^{2/3} \left[ \frac{dx^2}{1-x^2}+(1-x^2) \, d\phi^2 \right] \, .
\label{eq:umet}
\end{equation}
The unimodularity condition forces variations of the metric to be traceless, $g_{\mu\nu} \,\delta g^{\mu\nu}=0$.  It then follows that variation of the matter plus gravitational actions can only be forced to vanish up to a term of the form $\Lambda \,g_{\mu\nu}$.   The cosmological constant thus arises as a constant of integration rather than as a running parameter in the Lagrangian~\cite{Alvarez:2015sba}.  For a review of unimodular gravity, see, for example, Ref.~\cite{Alvarez:2023utn}.

From a quantum gravitational perspective, the restriction on $\sqrt{g}$ limits the possible graviton field configurations that contribute to the generating functional of correlation functions.   This implies that the running of the gravitational constant is not expected to be the same as in quantum Einstein gravity.   This has in fact been shown by direct calculation~\cite{Eichhorn:2013xr}; we summarize the main results relevant to unimodular gravity below.  

Let $g(k) \equiv k^2\, G(k)$, where $g(k)$ is the dimensionless gravitational constant and $k$ is the renormalization scale.  The renormalization
group equation takes the form $\partial _t g = \beta(g)$, where~\cite{Torres:2017ygl}
\begin{equation}
\beta(g) = 2 \,g \, \left(\frac{1-\omega\, g}{1- B \, g} \right) \, ,
\label{eq:beta}
\end{equation}
\begin{equation}
\omega = \frac{2 \,A_3-A_1}{2\, A_2} \, ,  \,\,\,\,\,\,\,\,\,\, B=\frac{A_3}{A_2} \,\,\,.
\label{eq:AB}
\end{equation}
Here $t=\ln k$, and the numerical values of the $A_i$ are determined from a functional renormalization group analysis~\cite{Eichhorn:2013xr}:
\begin{align}
A_1 &= 3\, (1300-309\sqrt{13}-325 \sqrt{17}) \\
A_2 &=936 \, \pi \\
A_3 &= 1625
\end{align}
Integrating Eq.~(\ref{eq:beta}), one finds that the combination
\begin{equation}
\frac{g(k)}{k^2\, (1-g(k)\, \omega)^{1-B/\omega}} = \mbox{ constant } 
\label{eq:rgeconst}
\end{equation}
is independent of the renormalization scale $k$.   Numerically, $1-B/\omega = 0.5158 \approx 1/2$; adopting this latter approximation, it follows
from Eq.~(\ref{eq:rgeconst}) that one may write
\begin{equation}
G(k) = \frac{1}{2} G_0\, \left(\sqrt{G_0^2\, k^4 \, \omega^2 + 4} - G_0\, k^2\, \omega\right) \,\,\, ,
\label{eq:Gofk}
\end{equation}
where $G_0 \equiv G(k=0)$.  Notice that $g(k)=k^2 G(k)$ approaches $1/\omega$ in the large $k$ limit, indicating the existence of an ultraviolet fixed point.  Eq.~(\ref{eq:Gofk}) agrees with the $G(k)$ presented in Ref.~\cite{Torres:2017ygl}; however, our identification of the renormalization scale will be entirely different.   The renormalization group improvement of the metric involves replacing $G(k)$ with $G(\rho)$ [or $G(r)$] by suitably identifying the renormalization scale $k$ with a length scale, $k\equiv{\cal K}(r)$.   It follows from Eq.~(\ref{eq:Gofk}) that
\begin{equation}
{\cal K}^2(\rho) =  \frac{1}{\omega} \frac{G_0^2 - G(\rho)^2}{G(\rho) \, G_0^2} \,\, .
\end{equation}
Since almost any choice of length scale is affected by the form of the metric, the question is whether one should use the classical metric or a quantum-corrected version that determines this scale self consistently.   If  ${\cal K}^2(\rho)$ is a function of $G(\rho)$ and its derivatives
 \begin{equation}
 {\cal K}^2(\rho) = F(G(\rho), G'(\rho), \ldots) \,\,\, ,
 \end{equation}
then our working hypothesis is that the quantum corrected $G(\rho)$ must satisfy
\begin{equation}
F(G(\rho), G'(\rho), \ldots) =  \frac{1}{\omega} \frac{G_0^2 - G(\rho)^2}{G(\rho) \, G_0^2} \,\,\,\,\,\,\,\,\,\, \mbox { (unimodular) } \,\,\ .
\label{eq:funi}
\end{equation}
One might compare this to the case of ordinary gravity using the corresponding renormalization group results given in Ref.~\cite{Platania:2019kyx}:
\begin{equation}
F_0(G(r), G'(r), \ldots) =  \frac{1}{\omega} \frac{G_0 - G(r)}{G (r) \, G_0} \,\,\,\,\,\,\,\,\,\, \mbox { (ordinary) } \,\,\ .
\label{eq:fein}
\end{equation}
Here, the prime refers to differentiation with respect to the argument, either $r$ or $\rho$, which is clear from context. Ref.~\cite{Platania:2019kyx} proposed a procedure in which the value of the gravitational coupling is iteratively improved, where the $n^{th}$ value ${\cal K}_n^2(r)$ is determined by the $(n-1)^{th}$ value $G_{n-1}(r)$.  In other words, the left-hand-side of Eq.~(\ref{eq:fein}) depends on $G_{n-1}(r)$, while the right-hand-side is a function of $G_n(r)$.   In this approach, the $G(r)$ in  (\ref{eq:fein}) would correspond to $G_\infty(r)$ in the notation of Ref.~\cite{Platania:2019kyx}.  However, in this recursive approach, $G_1(r)$ is an arbitrary seed function, and not every seed is guaranteed to reach the desired limit point.   For our study of the unimodular case, we instead constrain the quantum-corrected theory directly, postulating that $G(\rho)$ exactly solves Eq.~(\ref{eq:funi}), given a reasonable identification of the scale ${\cal K}(\rho)$.  We next turn to that issue.

\section{Two models} \label{sec:models}
In this section we consider two models that involve different types of identifications of the renormalization scale.  Of course, many other choices are possible, but we focus on two that are motivated by minimality and dimensional analysis.   The resulting metrics might be employed as benchmarks for comparison with other non-singular black hole metrics and for further phenomenological investigation.

\subsection{Model A: Identification involving derivatives of $G(\rho)$}
  
 As $G(\rho)$ is a coupling with dimensions, the logarithmic measure $\partial_\rho \ln [G(\rho)]$ provides perhaps the simplest identification of a characteristic  length scale that emerges in the quantum-corrected metric.   As an initial attempt, we assume the form
\begin{equation}
{\cal K}^2(\rho) = C_A \, \rho^{\,p} \, \left[ \frac{G'(\rho)}{G(\rho)}\right]^q \,\,\, ,
\label{eq:kform1}
\end{equation}
where $p$, $q$ and the constant $C_A$ are not yet determined.  We assume that $q$ is an integer, with $q \geq 1$.   This might be motivated by the observation that regular black hole metrics are not solutions to Einstein equations; new terms generated by the substitution of the corrected metric into the Einstein tensor that could be used to identify a physical length scale never depend on fractional or inverse powers of derivatives of $G(r)$.  As shown in Ref.~\cite{Platania:2019kyx}, objects like an effective energy density can depend on $G'(\rho) / G(\rho)$, though we comment later on why we don't pursue that type of identification.

From Eqs.~(\ref{eq:funi}) and (\ref{eq:kform1}), it follows that we need to solve
\begin{equation}
C_A \, \rho^{\,p} \, \left[ \frac{G'(\rho)}{G(\rho)}\right]^q  = \frac{1}{\omega} \frac{G_0^2 - G(\rho)^2}{G(\rho) \, G_0^2} \,\,\, .
\label{eq:solvethis}
\end{equation}
It is useful first to consider the asymptotic form of the solution for $G(\rho)$ at large $\rho$, which should approach the classical value $G_0$.  Writing
\begin{equation}
G(\rho) = G_0 [1 + \Delta(\rho)] \,\, , \,\,\,\,\,(\mbox{large } \rho)
\end{equation}
where $\Delta$ and its derivatives are assumed to be small for any acceptable solution, one finds
\begin{equation}
C_A \, \rho^{\,p}\, [\Delta'(\rho)]^q = - \frac{2}{\omega \, G_0} \Delta(\rho) + {\cal O}(\Delta^2) \,\,\, .
\label{eq:largerho}
\end{equation}
Consider first the case where $q>1$.  Integrating Eq.~(\ref{eq:largerho}) gives
\begin{equation}
\Delta^{1-\frac{1}{q}} = (1-1/q) \left(-\frac{2}{C_A\,\omega\, G_0}\right)^{1/q} \left\{\begin{array}{ccr} \ln\rho \, , & \mbox{   } &p=q \\ \frac{1}{1-p/q} \,\rho^{1-p/q}\, , &&p \neq q \end{array}\right.
\end{equation}
The requirement that $|\Delta|$ approaches $0$ as $\rho \rightarrow \infty$ requires that $p>q$.   However, at small radii, if we assume the 
asymptotic form 
\begin{equation}
G(\rho) = a\, \rho^s \,\,\, ,  \,\,\,\,\,(\mbox{small } \rho\,\, ,\mbox{ $a=$ a constant}) \,\,\, ,
\label{eq:aform}
\end{equation}
which can lead to regularity at the origin, substitution into Eq.~(\ref{eq:solvethis}) implies $s=q-p$.    The requirement that $p>q$ from the large $\rho$ asymptotic behavior implies $s$ is negative, which worsens the $\rho=0$ singularity.   For this reason, we exclude the possibility that $q>1$.

The case of $q=1$, however, is viable.   Repeating the analysis above, one finds
\begin{equation}
\Delta = \delta_A \exp\left[-\frac{2}{C_A \,\omega} \frac{1}{1-p} \rho^{1-p}\right] \,\,\, ,
\end{equation}
for some constant $\delta_A$.  Moreover, $s=1-p$ given our assumption that $G(\rho)$ vanishes as a power of $\rho$ near the origin.  Hence, for $s \geq 1$ which lead to singularity resolution, $\Delta \rightarrow 0$ as $\rho \rightarrow \infty$, as desired.  The case where $q=1$ allows us to integrate Eq.~(\ref{eq:solvethis}) and solve for $G(\rho)$ exactly:
\begin{equation}
G(\rho) = G_0 \tanh \left[\frac{1}{s} \frac{1}{\omega\, C_A G_0}\, \rho^s\right] \, .
\end{equation}
For definiteness, let us focus on the case of $s=1$, for which the metric is Eq.~(\ref{eq:umet}) with $f(\rho)$ replaced by the improved lapse function
\begin{equation}
f_I(\rho) = 1 - \frac{2 G_0 M}{(3 \rho)^{1/3}}\tanh \left[\rho/L^3\right] - \frac{\Lambda}{3} (3\rho)^{2/3} \,\,\, ,
\end{equation}
where the length scale $L$ is defined by $L^3 = \omega \,C_A G_0$.  This case is of interest since it corresponds to a de Sitter core, in which the constant $a=1/(\omega\, B)$ in Eq.~(\ref{eq:aform}), and
\begin{equation}
f_I(\rho) \approx 1 - \frac{2}{3^{1/3}} \frac{M}{\omega\, C_A} \, \rho^{2/3} + \cdots 
\label{eq:mAnear}
\end{equation}
near the origin, ignoring the $\Lambda$ term, which has the same form. If we require that $C_A$ is linear in $M$ so that the curvature at the origin is $M$-independent, and we take into account that $C_A$ has units of length, which is clear from the expression for $L^3$, this suggests that $C_A$ has the form
\begin{equation}
C_A = \hat{\xi}\, G_0 M \,\, ,
\end{equation}
for some constant $\hat{\xi}$.  This allows us to write
\begin{equation}
L^3 = \frac{1}{2} \, \hat{\xi} \, \omega \,r_s\, \ell_{Pl}^2. \,\,\, ,
\label{eq:Lcubed}
\end{equation}
where $r_s= 2 \,G_0 M$ is the classical Schwarzschild radius for $\Lambda=0$, and $\ell_{Pl}$ is the Planck length.   This gives some physical sense of the scale of the regulator $L$.

\subsection{Model B: Identification not involving derivatives of $G(\rho)$}

Here we consider the possibility that the scale  ${\cal K}(\rho)$ depends on $G(\rho)$, but not on its derivatives.  Since $G(\rho)$ appears in the improved metric always in the combination $G(\rho) M$, we consider combinations of this length scale and $\rho$ that could define a renormalization scale.   The most general form
is
\begin{equation} 
{\cal K}(\rho) = \frac{C_B}{r} F\left(\frac{G(\rho) M}{r}\right)\,\,\, \mbox{ with } r=(3 \, \rho)^{1/3} \,\, .
\label{eq:bscaleid}
\end{equation}
where $C_B$ is a dimensionless constant.   The function F is at this point arbitrary, so for definiteness we will consider the simple choice $F(x) = x^u$, for some power $u$.  Then, the self-consistency condition analogous to Eq.~(\ref{eq:solvethis}) is
\begin{equation}
\frac{C_B^2}{(3 \, \rho)^{2/3}}\left( \frac{G(\rho) M}{(3 \,\rho)^{1/3}}\right)^{2 u}
= \frac{1}{\omega} \, \frac{G_0^2 - G(\rho)^2}{G(\rho) \, G_0^2} \,\,\, .
\label{eq:solvethis2}
\end{equation}
We again assume the asymptotic form at small $\rho$ given by Eq.~(\ref{eq:aform}).  Eq.~(\ref{eq:solvethis2}) then implies
\begin{equation}
s=\frac{2}{3} \frac{1+u}{1+2 u} \,\,\,.
\end{equation}
Singularity resolution requires that $s \geq 1$, which leads to the following range for $u$:
\begin{equation}
-\frac{1}{2} < u \leq -\frac{1}{4} \,\, .
\end{equation}
We again focus on the case where we obtain a de Sitter core, namely $s=1$ and $u=-1/4$; this allows easy comparison the results of our first model and is a case where there is a clean, analytic solution to Eq.~(\ref{eq:solvethis2}) with a relatively simple form.   For this parameter choice, Eq.~(\ref{eq:solvethis2}) can be recast in the form
\begin{equation}
A(\rho)\,  y^{1/2} + y^2 -1 = 0 \,\,\,\, , \,\,\,\,\, A(\rho) \equiv \omega \,C_B^2 \sqrt{\frac{G_0}{3 M \rho}} \,\,\, ,
\end{equation}
where $y=G(\rho)/G_0$.  This has a unique, real solution
\begin{equation}
G(\rho) = \frac{G_0}{4} \left( \sqrt{\frac{2 \, A(\rho)}{\sqrt{z}}-z}-\sqrt{z}\right)^2 \,\,\, ,
\label{eq:bigG2}
\end{equation}
where
\begin{equation}
z=\sqrt[3]{\frac{A(\rho)^2}{2} + \sqrt{\frac{A(\rho)^4}{4} + \frac{64}{27}}}+\sqrt[3]{\frac{A(\rho)^2}{2} - \sqrt{\frac{A(\rho)^4}{4} + \frac{64}{27}}} \,\,\, .
\end{equation}
The renormalization group improved lapse function is $f_I(\rho) = 1 - 2\, G(\rho) M / (3 \rho)^{1/3} - \Lambda \, (3 \rho)^{2/3} / 3$, with $G(\rho)$ given by Eq.~(\ref{eq:bigG2}).   At small $\rho$ and again ignoring the $\Lambda$ term,
\begin{equation}
f_I(\rho)  \approx 1 - \frac{6}{3^{1/3}} \frac{M^2}{\omega^2 C_B^4} \, \rho^{2/3} + \cdots
\label{eq:mBnear}
\end{equation}
which can be compared to Eq.~(\ref{eq:mAnear}) in our first model.  From here, or from the expression for $A(\rho)$, we can again identify a characteristic 
length scale 
\begin{equation}
\tilde{L}^3 = \frac{\omega^2\, C_B^4 \,G_0}{3 M} \,\, ,
\end{equation}
analogous to $L^3$ of our first model.  Requiring that the curvature at the origin be independent of the mass $M$, the dimensionless constant must be proportional to $M^2$ and thus have the form
\begin{equation}
C_B^4 = \hat{\xi} \, G_0 \, M^2\,\,\, ,
\end{equation}
where $\hat{\xi}$ is a dimensionless constant that naively is of order one.  This gives
\begin{equation}
\tilde{L}^3 = \frac{1}{6} \, \hat{\xi} \, \omega^2 \,r_s\, \ell_{Pl}^2. \,\,\, ,
\end{equation}
which is similar to Eq.~(\ref{eq:Lcubed}), aside from a different power of $\omega$.

\begin{figure}[t]
\centering
\includegraphics[width=0.45\textwidth]{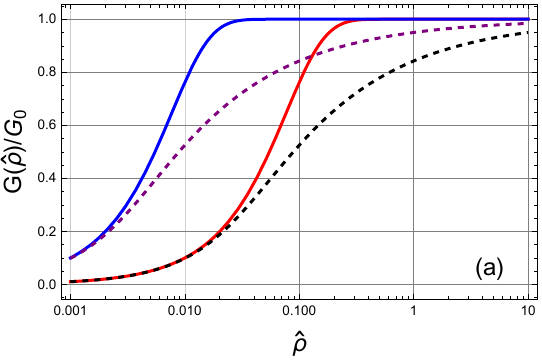} \hspace{1em}
\includegraphics[width=0.45\textwidth]{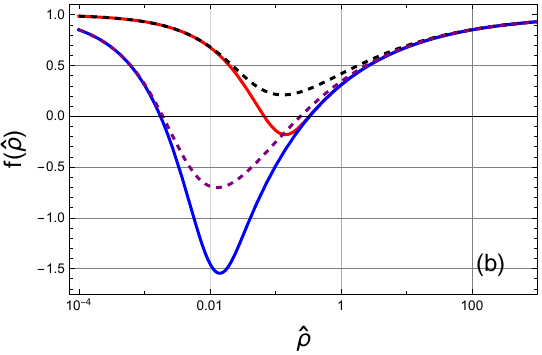}
\caption{{\bf (a)} $G(\hat{\rho}) / G_0$, where $\hat{\rho}\equiv \rho/r_s^3$, with $r_s \equiv 2 \, G_0 M$. The solid (dashed) curves correspond to Model A (Model B), for $r_s^3/L^3$ set to $10$ and $100$ (equating to very small black hole masses for illustrative purposes) with the larger ratio corresponding to the left-most curve. {\bf (b)}  The lapse function $f(\hat{\rho})$, following the same conventions and parameter choices as (a).}
\label{fig:fandg}
\end{figure}

For each of our models, we plot the running gravitational constant and the lapse function as a function of the dimensionless radial variable $\hat{\rho} \equiv \rho/r_s^3$, where $r_s = 2\, G_0 M$  in Fig.~\ref{fig:fandg}, for a few illustrative choices of $r_s/L$.   As expected, $G(\rho)$ goes to zero at the origin, necessary for the resolution of the singularity, and to $G_0$ at large radial distances, corresponding to the infrared limit of the renormalization group running.    The plots of the lapse functions are consistent with de Sitter cores [which is clear directly from Eqs.~(\ref{eq:mAnear}) and (\ref{eq:mBnear})] and asymptotically Minkowski spacetimes at large $\rho$.

The lapse functions show features that are familiar from other non-singular black hole models, namely mass ranges where there are two, one or no horizons.  The black hole mass gap refers to the range in mass $M$ where horizons exist.   To express this in notation familiar from the Hayward model, let us define a regulator $\ell$ by $L^3 \equiv 2\, G_0 M \, \ell^2$, and $\tilde{\ell}$ by an analogous relation involving $\tilde{L}$.   Numerically, we find for Model A that
\begin{equation}
G_0 \, M > 1.23 \, \ell \,\,\,\,\, \mbox{ with } \,\,\,\,\, \ell = \sqrt{\frac{\hat{\xi}\,  \omega}{2}}\,  \ell_{Pl}\,\, ,
\label{eq:gapA}
\end{equation} 
and for Model B
\begin{equation}
G_0 \, M > 2.25 \, \tilde{\ell} \,\,\,\,\, \mbox{ with } \,\,\,\,\, \tilde{\ell} = \sqrt{\frac{\hat{\xi}}{6}} \, \omega \, \ell_{Pl}\,\, ,
\label{eq:gapB}
\end{equation} 
where $\ell_{Pl}$ is the Planck length.   The inequalities in Eqs.~(\ref{eq:gapA}) and (\ref{eq:gapB})  are similar to those found in other nonsingular black hole models, as conveniently summarized in Table~1 of Ref.~\cite{Boos:2023icv}.   

\section{Discussion and Conclusions} \label{sec:conc}

In this work, we have considered the renormalization group improvement of unimodular black hole metrics assuming a self-consistent identification of the renormalization scale as a function of the radial variable.  Our choices were based on minimality and naive dimensional analysis.  While this solutions are by no means the only ones possible, they do provide interesting nonsingular unimodular black hole metrics that follow from a simple 
set of assumptions.   The scale identifications in our two models deserve further comment:

$\bullet$   In Model A, the scale was proportional to $G'(\rho)/G$, where the prime refers to the radial coordinate $\rho$.  This is perhaps the simplest approach to constructing a characteristic distance scale from a quantity that has nontrivial mass dimension and is a function of the radial variable. This is the same ratio that appears in an effective energy density that is used to identify a renormalization scale in the ordinary gravity case considered via the iterative approach of Ref.~\cite{Platania:2019kyx}.  One might wonder whether a stronger physical argument could be made in the unimodular black hole case by relating our choice to an energy density as well.  We have chosen not pursue this direction since the energy density in Ref.~\cite{Platania:2019kyx} appears in curvature invariants via terms that also depend on an effective equation of state parameter $w$, which itself is expressible in terms of first and second derivatives of $G$.  A naive identification of a characteristic physical scale must assume that these additional $w$-dependent factors are always of order unity, which may not  be the case for all values of the radial coordinate in an improved spacetime.   For this reason, we remain content with an identification motivated by minimality and dimensional considerations.  

$\bullet$ In Model B, we focused on the case where $u=-1/4$, that leads to a de Sitter core.  In this case, the renormalization scale identification used in Eq.~(\ref{eq:bscaleid}) may be written
\begin{equation}
{\cal K} \sim \frac{1}{ [G(\rho) M \rho \,]^{1/4}} \,\,\,\,\,\mbox{ with } \,\,\,\,\,  \rho  = \frac{1}{3} \left(\frac{A}{4\pi}\right)^{1/3} \,\,\, ,
\end{equation}
It is interesting that this scale is related to a spacetime volume: the product of a characteristic black hole time scale $G(\rho) M/c^3$ (with $c=1$)  and a spatial volume $\rho$ that is defined in terms of the invariant area of the SO(3) orbits of the metric, $A$.  This gives a geometric interpretation which is invariant under radial reparametrizations because it may be re-expressed entirely in terms on an area radius rather than the coordinate label $\rho$. Since it also depends on $M$ which is a global charge, one might characterize this choice as a spherically symmetric quasi-local setting of the renormalization scale.

The limitation of the identifications that we have studied is that they (like the one of Ref.~\cite{Platania:2019kyx}) are not fully diffeomorphism invariant.  We have not pursued scale identifications with this property, for example, ones involving curvature invariants~\cite{Held:2021vwd}, for the following reason: the right-hand sides of Eqs.~(\ref{eq:solvethis}) and (\ref{eq:solvethis2}) blow up as $\rho \rightarrow 0$ because $G(\rho)$ must vanish at the origin if we are to obtain a non-singular metric, which is our goal.   Curvature invariants constructed using the improved $G(\rho)$ are, by construction, non-singular.  Hence, identifying a suitable left-hand side of these equations using curvature invariants may be possible, but seems more likely to lead to an artificial-looking construction.  

The metrics we have discussed can be used in studies of black hole formation, thermodynamics and astrophysics, including  primordial black hole dark matter.  These topics have been considered in the context of other non-singular black hole models~\cite{Hayward:2005gi,Maluf:2018lyu,Bambi:2013ufa,Walia:2022ddq,Calza:2024xdh,Asmanoglu:2025agc}, and extending these studies to include the metrics presented here may be a worthy direction for future phenomenological study.

\begin{acknowledgments} 
CDC thanks the NSF for support under Grant No. PHY-2112460 and No. PHY-2411549.
\end{acknowledgments}


\end{document}